\documentstyle[twoside,fleqn,espcrc2]{article}

\newcommand{\AmS}{{\protect\the\textfont2
  A\kern-.1667em\lower.5ex\hbox{M}\kern-.125emS}}

\title{Which Higgs-Yukawa systems can possess non-trivial fixed points}

\author{Sergei V. Zenkin\address{Institute for Nuclear Research of
the Russian Academy of Sciences, 60th October Anniversary Prospect 7a,
117312 Moscow, Russia}
}

\begin{document}

\begin{abstract}
We argue that non-trivial fixed points bordering on the paramagnetic
and ferromagnetic phases are most likely to exist in the Higgs-Yukawa
systems that have a connected domain with the paramagnetic phase and
no ferrimagnetic phase. We find three examples of such systems; among
them is the U(1) system with naive fermions.
\end{abstract}

\maketitle

{\bf 1.} In this talk we want to emphasize the following three points:

(i) that the phase structure of the Higgs-Yukawa systems depend
crucially both on on their symmetry group and on the form of the
lattice fermion action, even though the coupling of the fermions to
the Higgs fields is the same;

(ii) that the most likely candidates
for the systems with non-trivial fixed points bordering on the
paramagnetic (PM) and ferromagnetic (FM) phases are the systems that
have a connected domain with the PM phase and no ferrimagnetic (FI)
phase;

(iii) that one of such systems is the U(1) system with naive
fermions.

We investigate the phase diagrams of the $Z_2$, U(1) and SU(2)
Higgs-Yukawa systems with radially frozen Higgs fields for three
types of chirally invariant lattice fermion actions. Our method is
based on the variational mean field approximation, where contribution
of the fermion determinant is calculated for weak and strong Yukawa
coupling regimes in a certain ladder approximation \cite{Z,TZ}. \\

{\bf 2.} The action of the system
\begin{eqnarray}
A = &-& 2\kappa \sum_{n, \mu, a} \phi^{a}_n
\phi^{a}_{n+\hat{\mu}} \cr
  &+& \sum_{m, n} \overline{\psi}_m (D_{mn}
+  y \tilde \Phi_m \delta_{mn})\psi_n
\end{eqnarray}
has two parameters: the scalar hopping parameter $\kappa$ and the
Yukawa coupling $y$; $\phi^a$ are real components of the group-valued
field $\Phi$, $\tilde \Phi = P_L \Phi^\dagger + P_R \Phi$. Given
group, the system is determined by the form of the lattice Dirac
operator $D$. At $\kappa > 0$ it corresponds to certain
continuum Higgs-Yukawa model; at non-positive $\kappa$ such a
correspondence is distroyed.

We shall require the system to be chirally invariant
and operator $D$ to have the form:
\begin{eqnarray}
D_{mn} &=& -D_{nm}, \cr
D_{mn} &=& \int_{p}
e^{i p (m - n)} i \, \gamma_{\mu} \, L_{\mu}(p),
\end{eqnarray}
with real
$L_{\mu}(p)$; $\int_p \equiv \int d^4 p / (2 \pi)^4$, $p_{\mu} \in (-\pi,
\pi)$.

We consider three examples of the lattice fermions satisfying these
conditions:

\noindent naive fermions with
\begin{equation}
L_{\mu}(p) = \sin p_{\mu};
\end{equation}
non-local SLAC fermions with
\begin{equation}
L_{\mu}(p) = p_{\mu};
\end{equation}
and fermions with
\begin{equation}
L_{\mu}(p) = \sin p_{\mu} \, \biggl[ 1 + \frac{(2 \sum_{\nu}
\sin^2 p_{\nu}/2)^2}
{\sum_{\nu} \sin^2 p_{\nu}}\biggr],
\end{equation}
whose action although is non-local, is
originated from the local mirror fermion action after integrating out
the mirror fermions (see \cite{MF} and also \cite{TZ}).

In all the cases the fermions couple to the Higgs fields in the same
way. \\

{\bf 3.} The method \cite{Z,TZ} yields closed analytical expressions
for the critical lines $\kappa (y)$ between the FM and PM phases and
between the antiferromagnetic (AM) and PM phases in the weak (W) and
strong (S) coupling regimes: $\kappa^W (y)$ and $\kappa^S (y)$,
respectively. Their explicit form for the $Z_2$, U(1) and SU(2)
systems can be found in \cite{TZ}. The important fact is that for
given group, the form of these line is mainly determined by four
constants: $G^W(0)$, $G^W(\pi)$, $G^S(0)$ and $G^S(\pi)$, which are
the values of the functions
\begin{eqnarray}
G^W (q) &=&  \int_p \frac{L(p) L(p+q)} {L^2 (p)
L^2(p+q)}, \cr
G^S (q) &=& \int_p L(p) L(p+q)
\end{eqnarray}
at $q = 0$ and $q = \pi$.
Moreover, in the cases of $Z_2$ and SU(2) these constants determine
singular points of the expressions for $\kappa^W (y)$ and $\kappa^S
(y)$, thereby determining the domains of the weak and strong coupling
regimes in these cases. These domains may be separated or they may
overlap. Both possibilities are realized in our examples.

As a result
we get three types of the phase diagrams \cite{TZ}:

(I) If these domains are
separated and FM-PM critical lines do not intersect AM-PM lines, we
have the diagram with two domains with PM phase separated by the
funnel with the FM phase at the intermediate values of $y$. This is
the case of $Z_2$ systems with the naive and mirror fermions.

(II) If
the FM-PM lines intersect AM-PM lines before possible intersection of
the lines $\kappa^W (y)$ and $\kappa^S (y)$, FI phase arises and PM
phase occupates two disconnected domains, too. This is the case for
the U(1) systems with more than one SLAC fermions or with mirror
fermions and for all the SU(2) systems.

(III) Finally, if the domains
of weak and strong coupling regimes overlap, or they are not
determined, like in the U(1) case, and the FI phase does not arise,
we have the diagram shown in Fig.~1. Domain with the PM phase is now
connected. This is the case for the U(1) systems with naive or one
SLAC fermions and for $Z_2$ system with SLAC fermions.

\begin{figure}[htb]
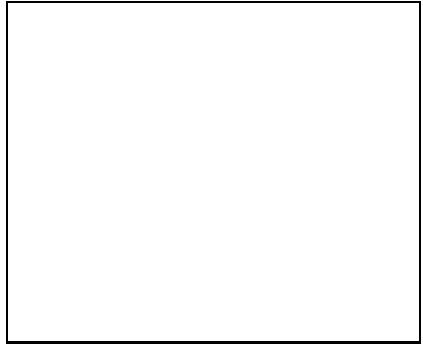

\vspace{9pt}
\framebox[55mm]{\rule[-21mm]{0mm}{43mm}}
\caption{Phase diagram of the type (III) (solid lines); fermion
condensate along
the FM-PM critical line (dashed lines).}
\label{fig:largenenough}
\end{figure}
\begin{figure}[htb]
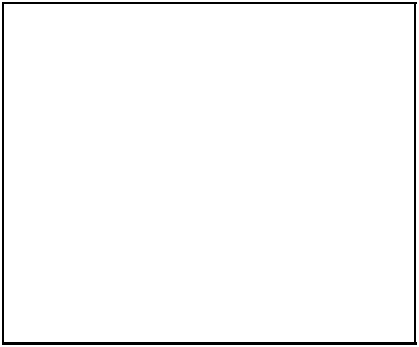

\vspace{9pt}
\framebox[55mm]{\rule[-21mm]{0mm}{43mm}}
\caption{Possible variant of the phase diagram of the type (III).}
\label{fig:toosmall}
\end{figure}

In all the cases fermion correlators have different properties in the
weak and strong coupling regimes, at least in the FM and PM phases,
thereby distinguishing FMW and FMS, and PMW and PMS phases \cite{Bo}
(see also \cite{TZ}). In particular, in PMW phase the fermions are
massless, while in PMS phase they are decoupled from the low lying
spectrum. \\

{\bf 4.} We now concentrate on the neighbourhood of the FM-PM
critical lines, since it is this domain is the most interesting for
application to the elementary particle physics. Consider the fermion
condensate along these lines. In our approximation this quantity has
the form
\begin{equation}
\langle \overline{\psi} \, \psi \rangle = - 2^{D/2} \, C(y)
\, \langle \phi \rangle + O(\langle \phi \rangle^2),
\end{equation}
where $\langle
\phi \rangle$ is v.e.v. of the Higgs field, and for given group
function $C(y)$ is determined in the weak and strong coupling regimes
by the constants $G^W(0)$ and $G^S(0)$, respectively, \cite{TZ}.

The important fact is that everywhere along the FM-PM critical lines
the condensate turnes out to be continuous smooth function of $y$,
except for the only point $A$ in the phase diagram of the type (III)
(see Fig.~1). In this figure dashed lines show the function $C(y)$
for the $U(1)$ system with naive fermions. The condensate can be
defined as the first order derivative of free energy of the system.
Therefore, its discontinuity in the point $A$ is an indication to a
first order phase transition separating FMW and FMS (and perhaps PMW
and PMS) phases. In Fig. 1 it is shown by the crossed line. Since the
condensate is an order parameter of the system, too, the point $A$
looks like tricritical point at which the first order phase
transition becomes a second order one.

Thus, we come to the scenario discussed in ref.~\cite{HN}. In view of
the results of ref.~\cite{Boc} that have given no evidence of non-trivial
behaviour of the SU(2) system in the point of forming the FI phase, the
point $A$ is the only candidate for non-trivial fixed point bordering
on the FM and PM phases. \\

{\bf 5.} So, we have found that only three of the considered systems
posses such points. Note, that all these points lie at
negative $\kappa$, where the systems have no their continuum
counterparts and are not reflection positive. Therefore, the
necessary condition for the systems to be of interest for continuum
physics is their physical positivitiy at $\kappa < 0$. However, if
this condition is satisfied and some of these points turn out to be
indeed non-trivial, the problem of universality of the Higgs-Yukawa
systems will arise. In any case, it would be interesting to find the
systems where such points are located at $\kappa > 0$.

Among these three systems is the U(1) system with the
naive fermions. The coordinates of the point $A$ in this case is:
$y^A \approx 1.34$, $\kappa^A \approx -0.43$ (for
two fermions). The
existence of such a point in this system, however, disagrees with
Monte Carlo (MC) results of ref.~\cite{HLN}, where FI phase has been
observed. We believe that this disagreement is due to the finite
lattice effects in the MC calculations. Our results for SU(2) system
with naive fermions and for U(1) and SU(2) systems with mirror
fermions are in agreement with MC results \cite{Bo,MF}. In these
cases the reasons for appearance of the FI phase are quite clear.
Therefore we see no reasons why our method should fail in this case.

The only change in the diagrams of the type (III) which we cannot
exclude is that the FM-PM critical line (and, perhaps, AM-PM line,
too) actually has a discontinuity at some value of $y$ not far from
$y^A$. In this case the point $A$ will be shifted and split into two
points, for example, as it is shown in Fig.~2. Investigation of this
question, however, requires different methods.\\

{\noindent \bf ACKNOWLEDGMENTS}\\

I am grateful to the International Science Foundation and to
Organizing Committee of the ``Lattice '94" for financial support, and
to Ph.~Boucaud and to M.~Stephanov for interesting discussions.

\end{document}